\documentclass{aastex61}
\received{26 April 2018}
\revised{8 July 2018}
\accepted{10 December 2018}
\submitjournal{AJ}

\shorttitle{Qatar-7b exoplanet}
\shortauthors{Alsubai et al.}

\usepackage{amsmath,texlogos}
\usepackage{natbib}
\usepackage{subfigure}
\usepackage{graphicx}
\usepackage{txfonts}
\usepackage[T1]{fontenc}
\usepackage{aecompl}
\usepackage{times}

\newcommand{\kms}{\ensuremath{\rm km\,s^{-1}}}
\newcommand{\ms}{\ensuremath{\rm m\,s^{-1}}}
\newcommand{\logg}{\ensuremath{\log{g}}}
\newcommand{\vsini}{\ensuremath{v\sin{i}}}
\newcommand{\vrot}{\ensuremath{v_{\rm rot}}}
\newcommand{\vlb}{\ensuremath{v_{\rm l.b.}}}
\newcommand{\Ha}{\ensuremath{\rm H\alpha}}

\newcommand{\feh}{\ensuremath{\left[{\rm Fe}/{\rm H}\right]}}
\newcommand{\mh}{\ensuremath{\left[m/{\rm H}\right]}}
\newcommand{\teff}{\ensuremath{T_{\rm eff}}}
\newcommand{\teq}{\ensuremath{T_{\rm eq}}}

\newcommand{\msun}{\ensuremath{{\,M_\odot}}}
\newcommand{\rsun}{\ensuremath{{\,R_\odot}}}
\newcommand{\lsun}{\ensuremath{{\,L_\odot}}}

\newcommand{\mstar}{\ensuremath{\,M_\star}}
\newcommand{\rstar}{\ensuremath{\,R_\star}}
\newcommand{\rhostar}{\ensuremath{\,\rho_\star}}
\newcommand{\mpl}{\ensuremath{\,M_{\rm P}}}
\newcommand{\rpl}{\ensuremath{\,R_{\rm P}}}
\newcommand{\mj}{\ensuremath{{\,M_{\rm J}}}}
\newcommand{\rj}{\ensuremath{{\,R_{\rm J}}}}

\newcommand\jk{\mbox{$J\!-\!K$}}

\begin{document}

\title{Qatar Exoplanet Survey: Qatar-7b -- A Very Hot Jupiter Orbiting a Metal Rich F-Star}

\correspondingauthor{Khalid Alsubai}
\email{kalsubai@qf.org.qa}

\author{Khalid Alsubai}
\affil{Hamad bin Khalifa University (HBKU), Qatar Foundation, PO Box 5825, Doha, Qatar}

\author{Zlatan I. Tsvetanov}
\affiliation{Hamad bin Khalifa University (HBKU), Qatar Foundation, PO Box 5825, Doha, Qatar}

\author{David W. Latham}
\affiliation{Harvard-Smitsonian Center for Astrophysics, 60 Garden Street,  Cambridge, MA 02138, USA}

\author{Allyson Bieryla}
\affiliation{Harvard-Smitsonian Center for Astrophysics, 60 Garden Street,  Cambridge, MA 02138, USA}

\author{Stylianos Pyrzas}
\affiliation{Hamad bin Khalifa University (HBKU), Qatar Foundation, PO Box 5825, Doha, Qatar}

\author{Dimitris Mislis}
\affiliation{Hamad bin Khalifa University (HBKU), Qatar Foundation, PO Box 5825, Doha, Qatar}

\author{Gilbert A. Esquerdo}
\affiliation{Harvard-Smitsonian Center for Astrophysics, 60 Garden Street,  Cambridge, MA 02138, USA}

\author{Ali Esamdin}
\affiliation{Xinjiang Astronomical Observatory (XAO), Chinese Academy of Sciences, 150 Science 1-Street, Urumqi, Xinjiang 830011, China }

\author{Jinzhong Liu}
\affiliation{Xinjiang Astronomical Observatory (XAO), Chinese Academy of Sciences, 150 Science 1-Street, Urumqi, Xinjiang 830011, China }

\author{Lu Ma}
\affiliation{Xinjiang Astronomical Observatory (XAO), Chinese Academy of Sciences, 150 Science 1-Street, Urumqi, Xinjiang 830011, China }

\author{Marc Bretton}
\affiliation{Observatoire des Baronnies Proven\c{c}ales (OBP), Le Mas des Gr\'{e}s, Route de Nyons, 05150 Moydans, France }

\author{Enric Pall\'{e}}
\affiliation{Instituo de Astrof\'{i}sica da Canarias (IAC), 38205 La Laguna, Tenerife, Spain}
\affiliation{Departamento de Astrof\'{i}sica, Universidad de La Laguna (ULL), 38206 La Laguna, Tenerife, Spain}

\author{Felipe Murgas}
\affiliation{Instituo de Astrof\'{i}sica da Canarias (IAC), 38205 La Laguna, Tenerife, Spain}
\affiliation{Departamento de Astrof\'{i}sica, Universidad de La Laguna (ULL), 38206 La Laguna, Tenerife, Spain}

\author{Nicolas P. E. Vilchez}
\affiliation{Hamad bin Khalifa University (HBKU), Qatar Foundation, PO Box 5825, Doha, Qatar}

\author{Timothy D. Morton}
\affiliation{Astronomy Department, University of Florida, Gainesville, FL 32611, USA}

\author{Hannu Parviainien}
\affiliation{Instituo de Astrof\'{i}sica da Canarias (IAC), 38205 La Laguna, Tenerife, Spain}
\affiliation{Departamento de Astrof\'{i}sica, Universidad de La Laguna (ULL), 38206 La Laguna, Tenerife, Spain}

\author{Pilar Monta\~{n}es-Rodriguez}
\affiliation{Instituo de Astrof\'{i}sica da Canarias (IAC), 38205 La Laguna, Tenerife, Spain}
\affiliation{Departamento de Astrof\'{i}sica, Universidad de La Laguna (ULL), 38206 La Laguna, Tenerife, Spain}

\author{Norio Narita}
\affiliation{Department of Astronomy, Graduate School of Science, The University of Tokyo, 7-3-1 Hongo, Bunkyo-ku, Tokyo 113-0033, Japan}
\affiliation{Astrobiology Center, National Institutes of Natural Sciences, 2-21-1 Osawa, Mitaka, Tokyo 181-8588, Japan}

\author{Akihiko Fukui}
\affiliation{Department of Astronomy, Graduate School of Science, The University of Tokyo, 7-3-1 Hongo, Bunkyo-ku, Tokyo 113-0033, Japan}
\affiliation{Astrobiology Center, National Institutes of Natural Sciences, 2-21-1 Osawa, Mitaka, Tokyo 181-8588, Japan}

\author{Nobuhiko Kusakabe}
\affiliation{Astrobiology Center, National Institutes of Natural Sciences, 2-21-1 Osawa, Mitaka, Tokyo 181-8588, Japan}

\author{Motohide Tamura}
\affiliation{Department of Astronomy, Graduate School of Science, The University of Tokyo, 7-3-1 Hongo, Bunkyo-ku, Tokyo 113-0033, Japan}
\affiliation{Astrobiology Center, National Institutes of Natural Sciences, 2-21-1 Osawa, Mitaka, Tokyo 181-8588, Japan}




\begin{abstract}

We present the discovery of Qatar-7b --- a very hot and inflated giant 
gas planet orbiting close its parent star. The host star is a relatively massive main 
sequence F-star with mass and radius $\mstar = 1.41 \pm 0.03 \msun$ and 
$\rstar = 1.56 \pm 0.02 \rsun$, respectively, at a distance $d = 726 \pm 26$ pc, 
and an estimated age $\sim1$ Gyr. With its orbital period of $P = 2.032$ days the 
planet is located less than 5 stellar radii from its host star and is heated to a high 
temperature $T_{\rm eq} \approx 2100$ K. From a global solution to the available 
photometric and radial velocity observations, we calculate the mass and radius of 
the planet to be \mpl\,=\,1.88$\pm$0.25\mj\ and \rpl\,=\,1.70$\pm0.03$\rj, 
respectively. The planet radius and equilibrium temperature put Qatar-7b in the 
top 6\%\ of the hottest and largest known exoplanets. With its large radius and 
high temperature Qatar-7b is a valuable addition to the short list of targets that 
offer the best opportunity for studying their atmospheres through transmission 
spectroscopy. 

\end{abstract}




\section{Introduction} \label{sec:Introduction}

While the detection of a habitable, Earth-like planet is undoubtedly seen as the "Holy Grail" 
of exoplanet research, hot Jupiters have enjoyed significant attention for many years and 
are still actively pursued. It is these gas giant planets, containing the bulk of the planetary 
system mass, that drive the shaping and final configuration of the planetary system. With 
growing numbers, statistical studies of populations became feasible, and intense efforts 
have been devoted into investigating possible correlations between the gas giants and the 
properties of their host stars, such as e.g., the now well-established positive correlation with 
the host's metallicity showing that hot Jupiters orbit stars with preferentially super-solar 
metallicities (e.g., \citealt{IdaLin04}, \citealt{FisherValenti05}, \citealt{Mordasini12}; see 
also \citealt{Maldonado13}, \citealt{Reffert15}, \citealt{Santos17}, \citealt{Buchhave2018}). 

Another important consideration in the framework of gas giant occurrence, is the role of the 
mass of the central star. Models of planet formation \citep[see e.g.,][]{IdaLin05, KennedyKenyon08} 
suggest a positive correlation, i.e., that the giant planet frequency increases with increasing 
stellar mass. While this prediction seems to be supported by observations in the solar-like, 
main-sequence (MS) FGK regime \citep[e.g.,][]{Cumming08}, it remains largely unvalidated 
in the MS-AF regime (with $\mstar \sim1.2-3.5 \msun$), mainly because of the technical 
difficulties in measuring Doppler radial velocity (RV) in MS-AF spectra, due to lack of 
spectral absorption lines and fast rotation.

To overcome this problem, GK-subgiants and giants have been used as proxys, based on 
the assumption that they are equally massive with- and descendants of MS-AF dwarfs 
\cite[e.g.][]{Johnson07, Johnson08, Johnson10a}, while having the advantages of being 
cooler (and therefore exhibiting more absorption lines) and less-rapidly rotating, allowing 
standard Doppler RV measurements. One of the main findings reported was an obvious 
paucity of gas giants at small separations ($\leq 1\,\rm{AU}$) around subgiant and giant 
stars, compared to solar-like dwarfs \cite[e.g.][]{Bowler10, Johnson10b, Reffert15}. Plausible 
mechanisms involving disk dispersal/depletion timescales versus Type-II migration 
\citep{Papaloizou07} timescale have been investigated (e.g. \citealt{BurkertIda07}, 
\citealt{Currie09}; see also \citealt{Ribas15}). We note however that some authors have 
challenged the validity of the assumptions regarding the mass of GK-subgiants and giants 
\citep{Lloyd11}, and proposed an alternative mechanism (albeit not necessarily mutually 
exclusive) involving orbital decay and tidal destruction \cite[e.g.][]{SchlWinn13,Villaver14}, 
sparking a debate on evolutionary models \citep{Johnson13,Lloyd13}.

In the above context, main-sequence AF stars can become targets of choice and provide 
crucial input into the impact of the host star's mass on the frequency and period distribution 
of gas giant planets, in the intermediate \mstar\ regime. Driven by a few transit-detections 
(OGLE2-TR-L9: \citealt{Snellen09}; WASP-33b: \citealt{collier10}; CoRoT-11b: 
\citealt{Gandolfi10}; KOI-13: \citealt{Rowe11}, \citealt{Mislis12}; KELT-9b: \citealt{Gaudi17}; 
KELT-20b: \citealt{Lund17}; MASCARA-2b: \citealt{Talens18}) and a dedicated RV survey 
\citep[][and the rest of the papers in the series]{Borgniet17}, close-in gas giants (hot Jupiters) 
around $\mstar \geq 1.5 \msun$ stars have started gaining in numbers, although their area 
of the parameter space remains underpopulated, particularly in the short period 
($P \leq 5\,\rm{d}$) regime.

In this article we present the discovery of Qatar-7b --- a (very) hot Jupiter transiting a 
fast rotating F-star --- found by the Qatar Exoplanet Survey (QES) The structure of the 
paper is as follows: in Section \ref{sec:Observations} we describe the observations -- both 
the discovery photometry and the follow-up spectroscopy and photometry. In Section 
\ref{sec:Analysis} we discus the data analysis and the global system solution using 
simultaneous fits of the RV and transit light curves. The results are summarized in 
Section \ref{sec:conclusions}. 

\section{Observations} \label{sec:Observations} 

\subsection{Discovery photometry} \label{subsec:DiscPhot}

The survey data were collected with QES, hosted by the 
New Mexico Skies Observatory\footnote{http://www.nmskies.com}, Mayhill, NM, 
USA. A detailed description of QES can be found in our previous publications, e.g., 
\cite{alsubai2013}, \cite{alsubai2017}. In brief, QES uses two overlapping wide field 
135\,mm (f/2.0) and 200\,mm (f/2.0) telephoto lenses, together with four 400\,mm (f/2.8) 
telephoto lenses, mosaiced to image an $11{\degr}\,\times\,11{\degr}$ field on the sky. 
Each lens is equipped with a FLI ProLine PL16801 camera utilizing a KAF-16801E 
4k$\times$4k detector. 

The discovery light curve of Qatar-7b contains $\sim$11\,000 data points obtained with 
one of the 400\,mm lenses from October 2012 to December 2014. The data were reduced 
with the QES pipeline, which performs bias-correction, dark-current subtraction and 
flat-fielding in the standard fashion, while photometric measurements are extracted using 
the image subtraction algorithm by \cite{dbdia}; a more detailed description of the pipeline 
is given in \cite{alsubai2013}.

The extracted light curves are detrended using a combination of the Trend Filtering 
Algorithm (TFA, \citealt{kovacs1}) and the Doha algorithm \citep{mislis}. Qatar-7b was 
identified as a strong candidate during a search for transit-like events using the Box 
Least Squares algorithm (BLS, \citealt{kovacs2}), following a procedure similar to that 
described in \cite{collier}. Note that, although the initial candidate selection is an 
automatic procedure, the final vetting is done by eye. 

\subsection{Follow-up photometry} \label{subsec:FollowPhot}

Follow-up photometric observations of several transits of Qatar-7b were collected 
at five different observatories with the following combination of telescopes and 
instruments: 
(a) {FLWO:} we used the 1.2\,m telescope at the Fred L.\,Whipple Observatory 
(Mount Hopkins, Arizona), together with KeplerCam, a 4K $\times$ 4K CCD, with 
an on-sky FOV of 23$^{\prime} \times 23^{\prime}$; 
(b) {QFT:} we used the 0.5\,m Qatar Follow-Up Telescope (New Mexico Skies 
Observatory, Mayhill, New Mexico), equipped with a 1k$\times$1k Andor iKon-M 
934 CCD, yielding a FOV of 13$^{\prime} \times 13^{\prime}$; 
(c) {OBP:} we used the 0.82\,m telescope of the Observatoire des Baronnies 
Proven\c{c}ales\footnote{http://www.obs-bp.fr} (Provence-Alpes-C\^{o}te d'Azur, 
France), equipped with an FLI ProLine PL230 CCD, with a 2k$\times$2k E2V 
detector, resulting in a 23$^{\prime} \times 23^{\prime}$ FOV; and 
(d) {XAO:} we used the Nanshan One-meter Wide-field Telescope (NOWT) of 
the Xinjiang Astronomical Observatory (XAO) equipped with a 4k$\times$4k 
CCD with full on-sky FOV of $1.28\arcdeg \times 1.28\arcdeg$. To shorten the 
readout time and increase the duty cycle only a $1200 \times 1200$ pixels 
subframe was used resulting in a 22.5$^{\prime} \times 22.5^{\prime}$ area 
on the sky. 
(e) {TCS:} we used the 1.52\,m Telescopio Carlos Sanchez (TCS) at the Teide 
Observatory (Tenerife, Canary Islands), and MuSCAT2 instrument, which takes 
images in 4 filters simultaneously. Each channel is equipped with a 1K $\times$ 1K 
CCD, resulting in a 7.4$^{\prime} \times 7.4^{\prime}$ on-sky FOV. For a detailed 
description of MuSCAT2 and its dedicated photometric pipeline see \cite{MuSCAT2}.

\begin{figure*}
\centering
\includegraphics[width=0.45\textwidth]{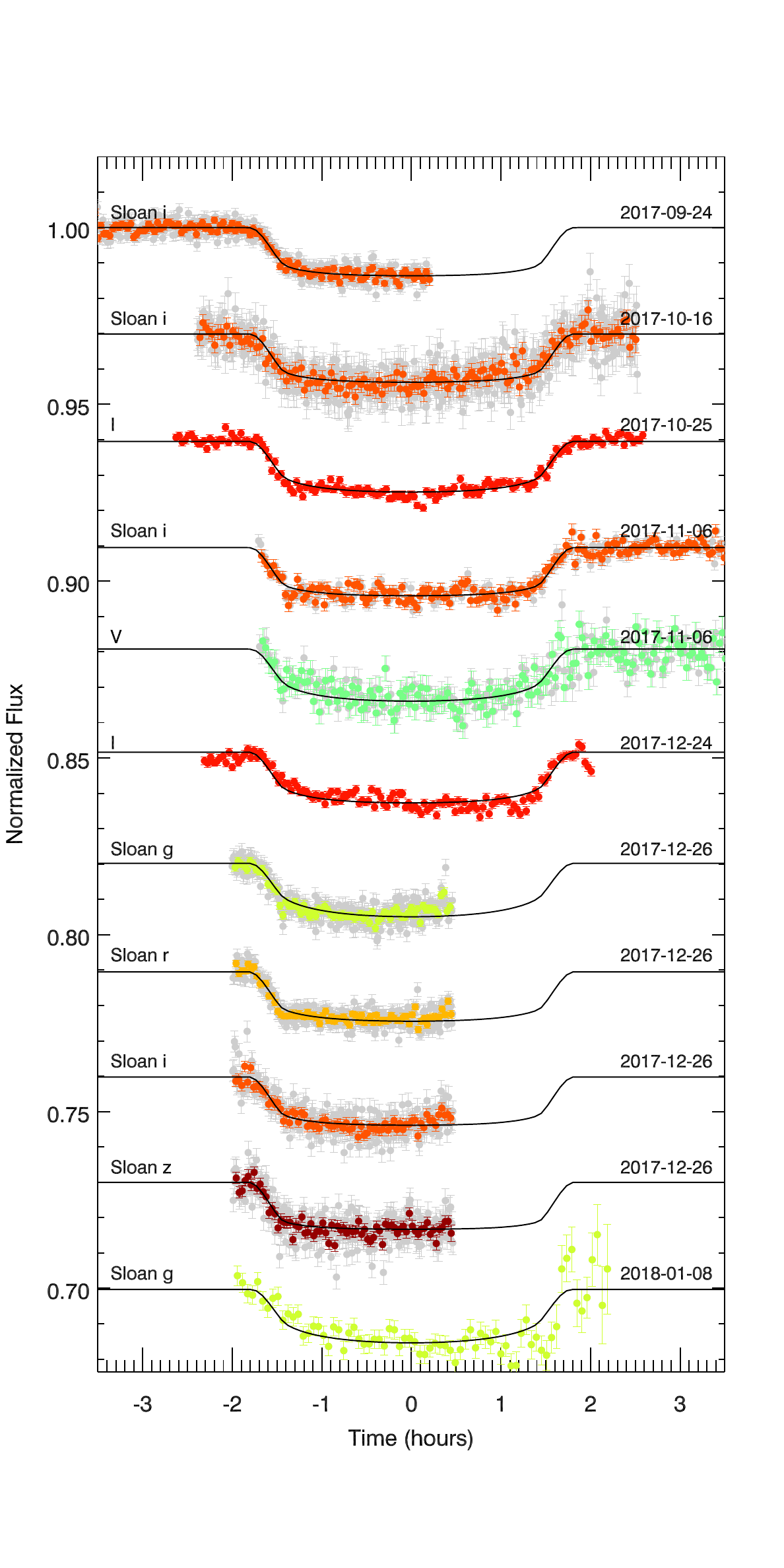}
\includegraphics[width=0.45\textwidth]{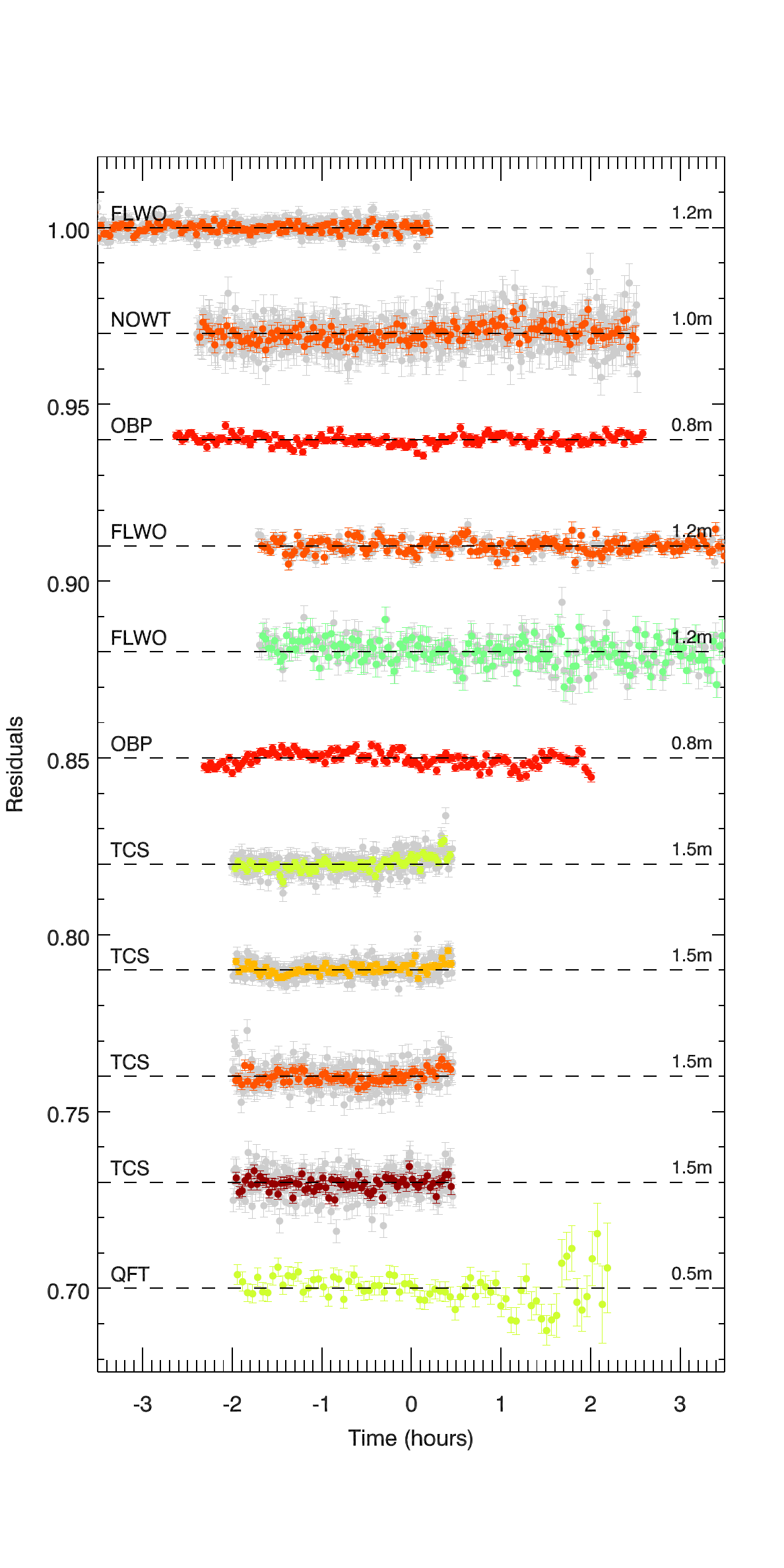}
\caption{The eleven follow-up transit light curves of Qatar-7b. In the {\it left} panel, the 
light curves are ordered from top to bottom as they appear in Table\,\ref{table:ObsLog} 
and have been shifted vertically for clarity. The solid, black lines represent the best 
model fit (see Section\,\ref{subsec:EXOFAST}). The residuals from the fits are shown 
in the {\it right} panel. The individual data points are color coded according to the 
filter used and for observations taken at FLWO, XAO and Teide observatories we 
show both the original data points (light gray) as well as the data binned to a 
uniform cadence of 2 min, while for observation taken at OBP and QFT we only 
show the original data points as thier cadence is $\geq 2$ min. The filter, date of 
observation, observatory and telescope size are also given in the two panels.}
\label{fig:q7bTR}
\end{figure*}

A summary of our follow-up photometric observations is given in Table 
\ref{table:ObsLog}. The resulting light curves, along with the best model fit and 
the corresponding residuals, are plotted in Figure\,\ref{fig:q7bTR}.

\begin{table}
\centering
\caption{Log of follow-up transit observations for Qatar-7b. See text for details on 
telescopes/instruments.}
\label{table:ObsLog}
\begin{tabular}{ccccccc}
\hline\hline
Obs ID & Date & Telescope & Filter & Cadence, s \\
\hline
1 & 2017-09-24 &  FLWO & i & 35  \\
2 & 2017-10-16 &  NOWT & i & 30  \\
3 & 2017-10-24 &  OBP & I & 125  \\
4 & 2017-11-06 &  FLWO & i & 84  \\
5 & 2017-11-06 &  FLWO & V & 84 \\
6 & 2017-12-24 &  OBP & I & 125  \\
7 & 2017-12-26 &  TCS & g & 30  \\
8 & 2017-12-26 &  TCS & r & 30  \\
9 & 2017-12-26 &  TCS & i & 30  \\
10 & 2017-12-26 & TCS & z & 30  \\
11 & 2018-01-08 &  QFT & g & 204  \\
\hline
\end{tabular}
\end{table}

\subsection{Follow-up spectroscopy} \label{subsec:FollowSpec}

Follow-up spectroscopic observations for Qatar-7b were obtained in the same 
manner as for all other QES candidates. We used the Tillinghast Reflector Echelle 
Spectrograph (TRES) on the 1.5\,m Tillinghast Reflector at FLWO. The light from 
the object was fed to the spectrograph through the medium fiber, resulting in 
a resolving power of $R \sim$ 44,000 and a velocity resolution element of 6.8 \kms\ 
FWHM. Two exposures of a Th-Ar hollow-cathode lamp illuminating the science 
fiber were taken immediately before and after each each science spectrum to 
establish precise wavelength calibration.

\begin{table}
\centering
\caption{Relative RVs and BS variations for Qatar-7.}
\label{table:RV}
\begin{tabular}{ccc}
\hline\hline
BJD$_{TDB}$       &RV (\ms)          &BS (\ms)       \\
\hline
$ 2457993.94705 $ & $366.9 \pm 139.5 $ & $ -9.0 \pm 65.0 $ \\
$ 2457994.91573 $ & $-271.6 \pm 77.2 $ & $ 129.0 \pm 59.3 $ \\
$ 2458034.63495 $ & $244.2 \pm 81.5 $ & $ 18.7 \pm 50.8 $ \\
$ 2458035.61646 $ & $-109.1 \pm 97.8 $ & $ -7.1 \pm 48.7 $ \\
$ 2458037.80307 $ & $-2.2 \pm 137.0 $ & $ -33.5 \pm 65.6 $ \\
$ 2458039.68421 $ & $-291.6 \pm 93.9 $ & $ 70.2 \pm 30.8 $ \\
$ 2458040.60790 $ & $125.6 \pm 109.4 $ & $ -21.1 \pm 57.3 $ \\
$ 2458042.69416 $ & $206.3 \pm 81.1 $ & $ -22.6 \pm 68.9 $ \\
$ 2458095.60236 $ & $432.7 \pm 123.5 $ & $ -97.6 \pm 77.1 $ \\
$ 2458107.71464 $ & $457.4 \pm 173.3 $ & $ 144.9 \pm 85.8 $ \\
$ 2458116.58480 $ & $-116.6 \pm 131.6 $ & $ -135.2 \pm 102.0 $ \\
$ 2458127.59278 $ & $-0.0 \pm 123.5 $ & $ -14.3 \pm 54.1 $ \\
$ 2458156.61948 $ & $458.9 \pm 180.8 $ & $ -16.2 \pm 57.8 $ \\
$ 2458157.61827 $ & $-99.5 \pm 130.1 $ & $ -6.4 \pm 97.6 $ \\
\hline
\end{tabular}
\end{table}

\begin{figure}
\centering
\includegraphics[width=8.5cm]{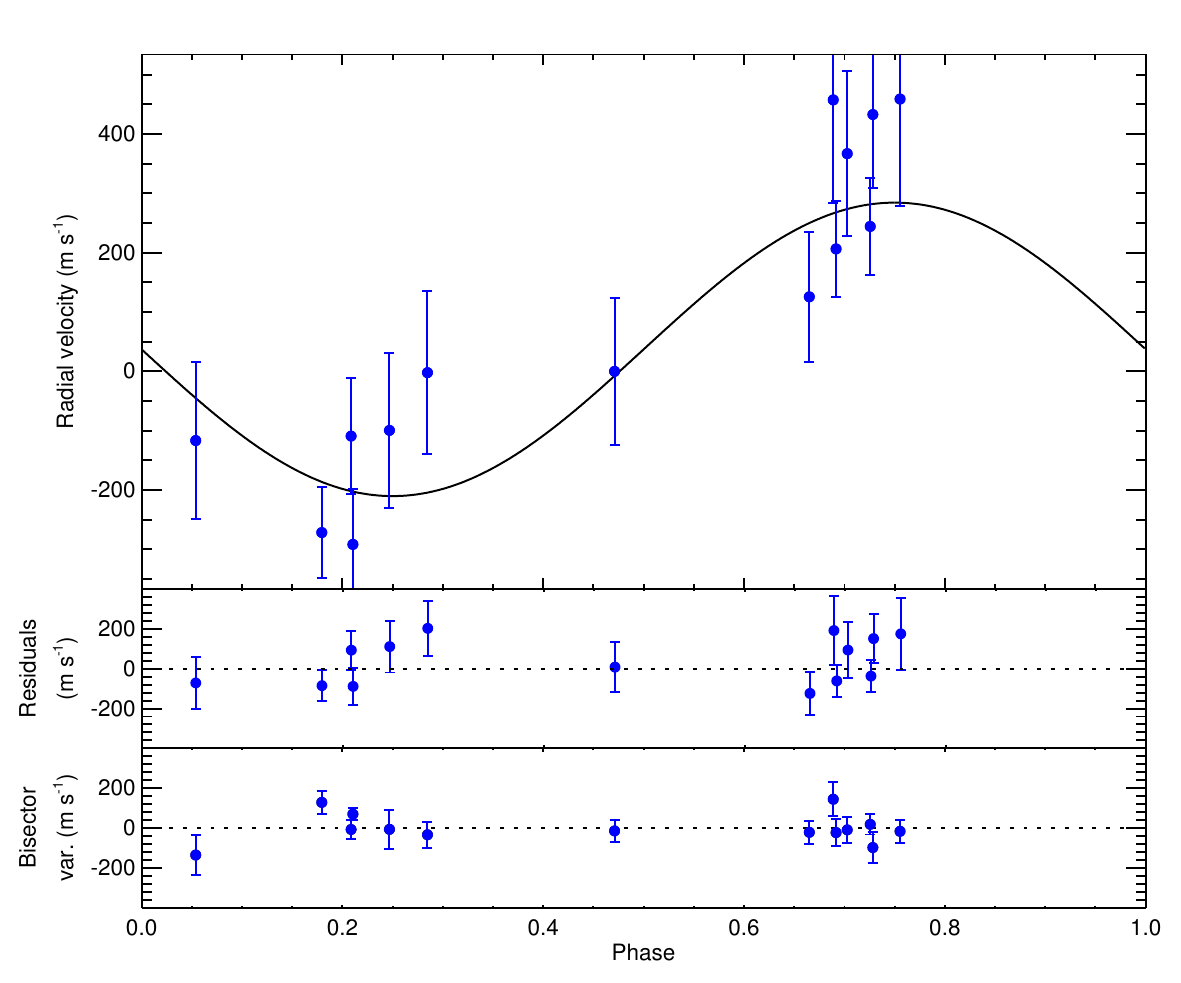}
\caption{Orbital solution for Qatar-7b, showing the velocity curve and observed 
velocities and the bisector values.}
\label{figure:q7bRV}
\end{figure}

For Qatar-7 we obtained 14 spectra between August 28, 2017 -- February 8, 2018 
with exposure times in the range 30--60 min resulting in an average signal-to-noise 
ratio per resolution element (SNRe) of $\sim$25 at the peak of the continuum close 
to the Mg b triplet near 519 nm. 
We selected the best spectrum as our reference one, and measured relative radial 
velocities (RV) using the cross-correlation technique; under this framework, a specific 
set of echelle orders is chosen, meeting the following two criteria (i) good SNRe and 
(ii) minimal telluric lines contamination. All the 
observed spectra subsequently undergo order by order cross-correlation against the 
reference spectrum. By default, the RV value of the reference spectrum is zero, while 
the error assigned to this value is simply the median of all other errors of our RV 
measurements. In order to exclude other astrophysical phenomena, that could 
potentially mimic the orbital-motion induced periodic signal evident in our RV 
measurements, we also calculated the line profile bisector spans (BS). The full process 
of measuring the RVs and BSs is described in detail in \citet{buchhave2010}. We collect 
the RV and BS values in Table \ref{table:RV} and plot them in Figure \ref{figure:q7bRV}.

\section{Analysis and Results} \label{sec:Analysis} 

\subsection{The host star} \label{subsec:HostStar}

The host star of Qatar-7b (2MASS\,J23540364+3701185, henceforth Qatar-7) is a relatively 
faint, $V$\,=\,13.03 mag, mid-to-late F-type star\footnote{An estimate from a multi-color fit 
to the $V$, $J$, $H$ and $K$ magnitudes, using a standard Random-Forest classification 
algorithm}. In what follows, we present our analysis of the host star, while 
Table\,\ref{table:StarPars} presents a comprehensive summary.

\subsubsection{Spectroscopic parameters} \label{subsec:SpecPars}

To determine the host star atmospheric characteristics -- effective temperature (\teff), 
surface gravity (\logg), and metallicity (\feh), as well as the projected rotational 
velocity (\vrot) -- we analyzed the available spectroscopic data used for the RV 
measurements (see Section\,\ref{subsec:FollowSpec}) through the Stellar Parameter 
Classification tool (SPC, \citealt{SPC})). Because individual TRES spectra have 
relatively low S/N ratio, all 14 spectra were wavelength shifted and co-added to 
form a high signal-to-noise spectrum representative of the host star. 

The SPC determines the stellar parameters from a multi-dimentional surface fit to 
the cross correlation of the observed spectrum with a library of Kurucz synthetic 
spectra (\citealt{ODF}). The SPC determined values are: $\teff\,=6311 \pm\ 50$ K, 
$\logg\ = 4.26 \pm\ 0.10$, $\feh =0.29 \pm 0.08$, and line broadening 
$\vlb = 14.7 \pm\ 0.5$ \kms.

\begin{figure}
\centering
\includegraphics[width=8.8cm]{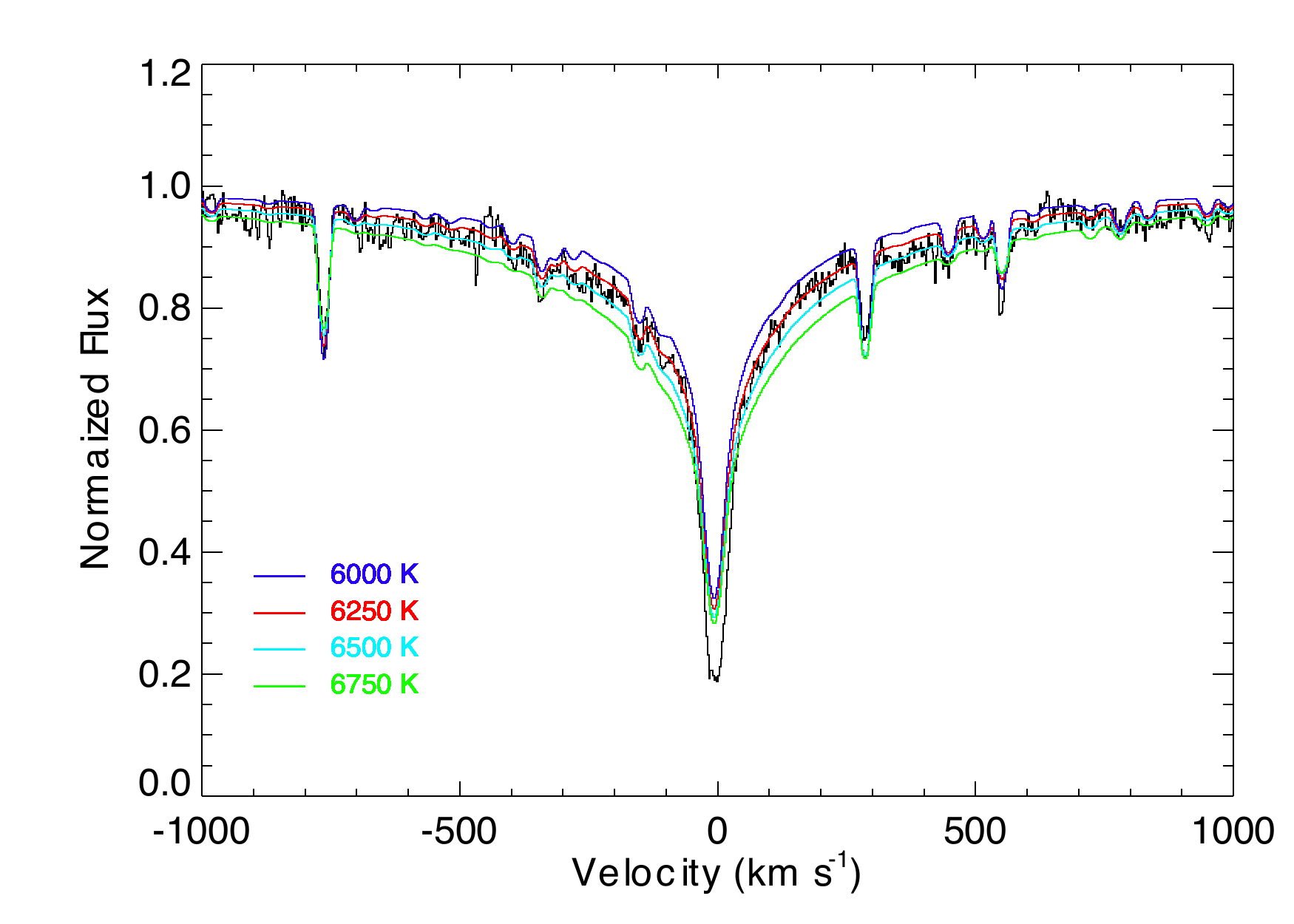}
\caption{H$\alpha$ profile from the combined spectrum of Qatar-7 compared with 
four synthetic spectra with \feh\,=0.25, \logg\,=4.2 and \teff\,=6000, 6250, 6500, and 
6750 K, respectively.}
\label{fig:HaProfile}
\end{figure}

In solar type main sequence stars the wings of the \Ha\ line profile have strong 
sensitivity to temperature variations and at the same time relatively weak sensitivity 
to changes in the surface gravity and metallicity, which makes this feature a good 
temperature indicator (e.g., \citealt{Fuhrmann93}; \citealt{Barklem02}). As a 
consistency check on \teff\ we compared the observed \Ha\ profile outside the 
core in the co-added spectrum to sythetic spectra generated with the 
Spectroscopy Made Easy package (SME, \citealt{SME})
using the Kurucz ATLAS9 grid of models. Template spectra were generated with 
\logg\,=\,4.2, \feh\,=\,0.25 and \teff\,=\,6000, 6250, 6500, and 6750 K. In Figure 
\ref{fig:HaProfile}, in velocity space, we show the observed \Ha\ profile and 
the four synthetic spectra. It is clear the \Ha\ line profile is best reproduced by 
synthetic spectra with \teff\ in the range 6250-6500 K. We note here, that the 
shape of the observed \Ha\ profile is sensitive to the continuum normalization, 
which is difficult to pinpoint for such a broad line in an echelle spectrum. For that 
reason, we consider this comparison only as a rough check on \teff, yet, it agrees 
with our previous estimates.

\subsubsection{SED fit} \label{subsec:SEDfit}
 
Qatar-7 is detected by most of the large-scale surveys and, as such, photometric 
measurements are available across the spectrum, from the NUV (GALEX) to the 
mid-IR (WISE). These measurements are gathered in Table\,\ref{table:StarPars}. 
We note that for the Sloan $u,g,r,i,z$ bands, the \emph{actual} SDSS 
photometry\footnote{http://skyserver.sdss.org/dr14} is classified as ``unreliable'', 
accompanied by a variety of cautionary flags primarily related to saturated and 
interpolated pixels. To ameliorate the problem, we turned to the APASS catalogue 
\citep{apass} and the work of \cite{Pickles2010} (PD10) and adopt the $g,r,i$ 
values from APASS and the $u,z$ values from PD10.

\begin{figure}
\centering
\includegraphics[width=8.8cm]{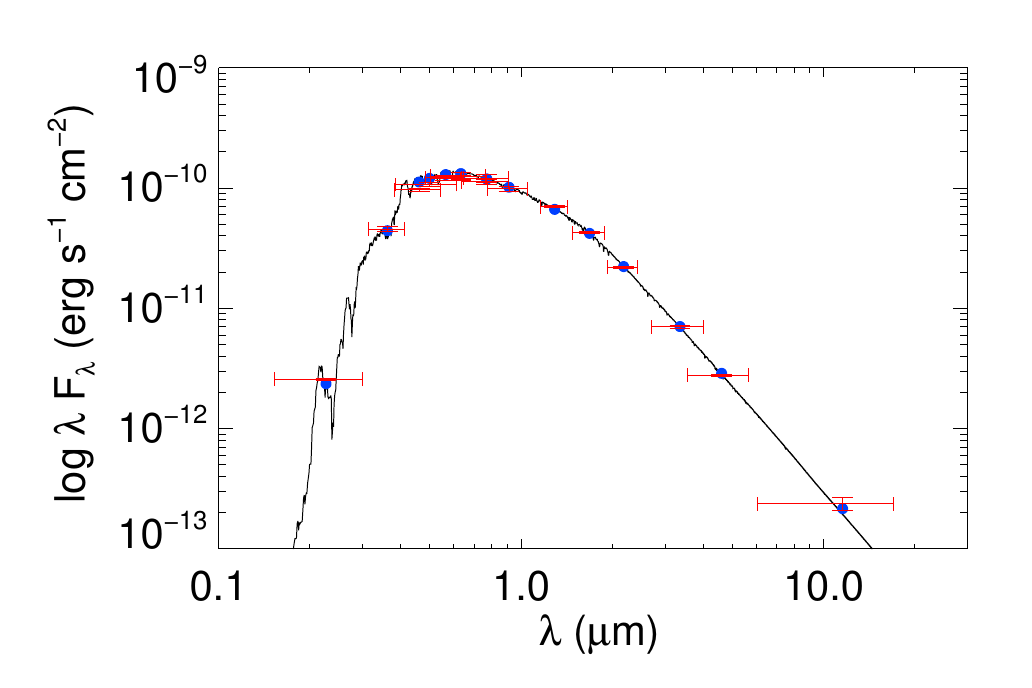}
\caption{Spectral Energy Distribution (SED) fit for Qatar-7b host star. The 1$\sigma$ 
uncertainties of the photometric measurements (Table\,\ref{table:StarPars}) are 
plotted as vertical bars, and the effiective width of the passbands is shown by  
horizontal bars. The solid curve is the best fit model SED where stellar parameters 
\rstar, \teff, \logg, and \feh\ were kept fixed at the values derived from the global fit 
(Table\,\ref{table:GlobalFit}). }
\label{fig:SEDfit}
\end{figure}

We used the broadband measurements combined with the distance measured by 
{\it Gaia} to fit a SED using the NextGen library of theoretical models and solve for 
the extinction and stellar radius \rstar. We note that the shape of the SED is 
determined by the \teff\ and extinction, while the value of \rstar\ is dependent 
on the distance. In fitting the SED we imposed a Gaussian prior of 0.05 on the 
extinction mean value of $A_{V} \approx 0.35$ mag as suggested by the Galactic 
dust reddening maps (\citealt{Extinction}). Figure \ref{fig:SEDfit} shows that the 
SED of a $\teff \sim 6400$ K main sequence star fits very well the broadband 
photometry for a distance $d = 720$ pc and extinction $A_{V} = 0.36$ mag 
(uncertainties are given in Table \ref{table:StarPars}). 

\subsubsection{Age} \label{subsec:HostAge} 

To obtain an estimate of the age of Qatar-7, we used three different age 
indicators---gyrochronology, isochrone fitting, and lithium abundance. First, the 
star is a relatively fast rotator, which is indicative of a young age. For the 
gyrochronology method we applied the rotational period--color--age relations, as 
calibrated by \cite{brown}, assuming the stellar rotation axis is perpendicular to the 
orbital plane of the planet. The stellar projected equatorial rotational velocity, \vsini, 
was estimated form the line broadening, \vlb, measured by SPC and accounting for the 
macroturbulence velocity, $v_{\rm mac}$, contribution. We esitamted $v_{\rm mac} = 
5$ \kms\ using the calibrations of \cite{Melendez2012} (average of their Eqs.\ (E.1) 
and (E.2) valid for the \teff\ of our star) and assuming $v_{\rm mac}^{\sun} = 3.6$ \kms. 
As a result we adopted $\vrot \equiv \vsini = (\vlb^{2} - v_{\rm mac}^{2})^{1/2} = 13.8$ \kms\ 
for the rotational velocity and we used the reddening corrected values for the 
\bv\ and \jk\ colors. The gyrochronology estimated age is $\tau_{\rm gyr} = 0.5 \pm 0.1$ 
Gyr, where the uncertainty reflects only the errors in color measurements and 
determining the stellar rotation period.

An alternative estimate of the host star age can be obtained from comparing the
observationally determined parameters \teff\ and \rhostar\ with theoretical evolution 
models. Stellar effective temperature, \teff, is determined form the spectroscopic 
analysis (Section \ref{subsec:SpecPars}) while stellar density, \rhostar, is derived from 
the transit light curves analysis. We note that the fit to the transit light curves is obtained 
by varying three parameters --- the relative planet radius, $\rpl/R_{\star}$, the impact 
parameter, $b = a\cos{i}/\rstar$, and the relative semi-major axis, $a/\rstar$ (see 
e.g., \citealt{MandelAgol02}, \citealt{Pont07}). Using the Kepler's third law and 
assuming $\mstar >> \mpl$ and a circular orbit the density of the star is directly 
determined by $a/\rstar$ and the orbital period (e.g., \citealt{Seager03}). 

\begin{figure}
\centering
\includegraphics[width=8.8cm]{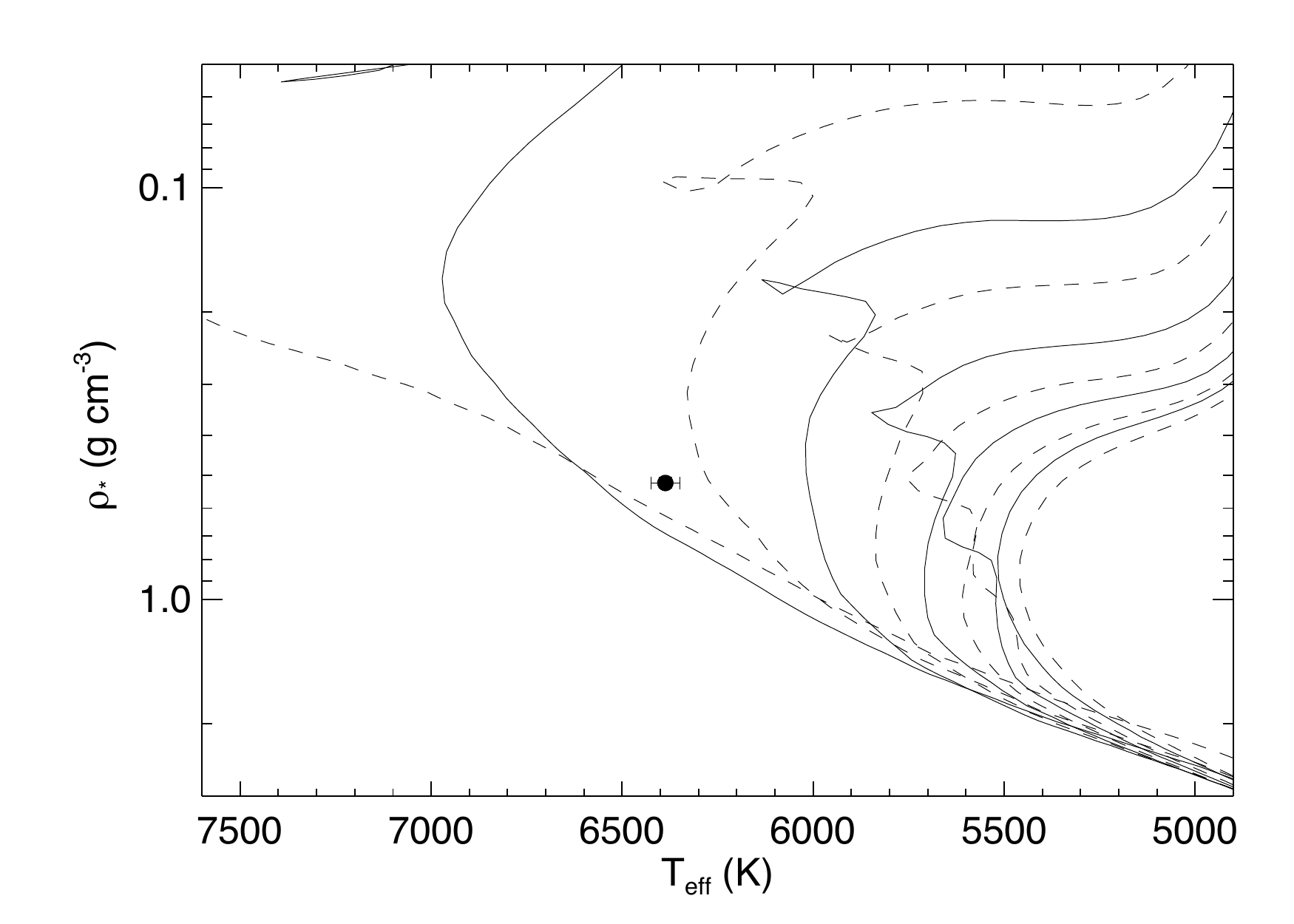}
\caption{Model isochrones from the YY series (\citealt{YY}) for the measured 
metallicity of Qatar-7b. Isochrones for ages 0.6 Gyr and 1.0--10.0 Gyr in 1 Gyr 
intervals are plotted in alternating dashed and solid lines. Ages are increasing left 
to right.}
\label{fig:Teff_rho_YY}
\end{figure}

In Figure \ref{fig:Teff_rho_YY} we compare the stellar density \rhostar\ and \teff\ with 
model isochrones from the \cite{YY} database. The isochrones plotted are calculated 
for the metallicity suggested by the spectroscopic analysis, $\mh\,= 0.25 \pm 0.10$, 
and cover a wide range in age (0.6-12 Gyr). The best match to the observationally 
measured parameters \rhostar, \teff\ is achieved for stellar age $\tau\ = 1.0 \pm 0.5$ 
Gyr, where the uncertainty is driven by the observational errors. 

In the course of the stellar evolution Li is destroyed at temperatures $\sim 2.5\times10^{6}$ 
K when it is transported to the innermost layers of the star through convective motions. 
Although the level of Li burning is controlled by several factors, e.g., convective zone 
thickness, stellar mass, and metallicity,  the degree of Li depletion in stars depends 
primarily on the stellar age in the sense that older stars have lower Li abundance 
(e.g., \citealt{Melendez2014}, \citealt{TucciMaia2015}). 

\begin{figure}
\centering
\includegraphics[width=8.8cm]{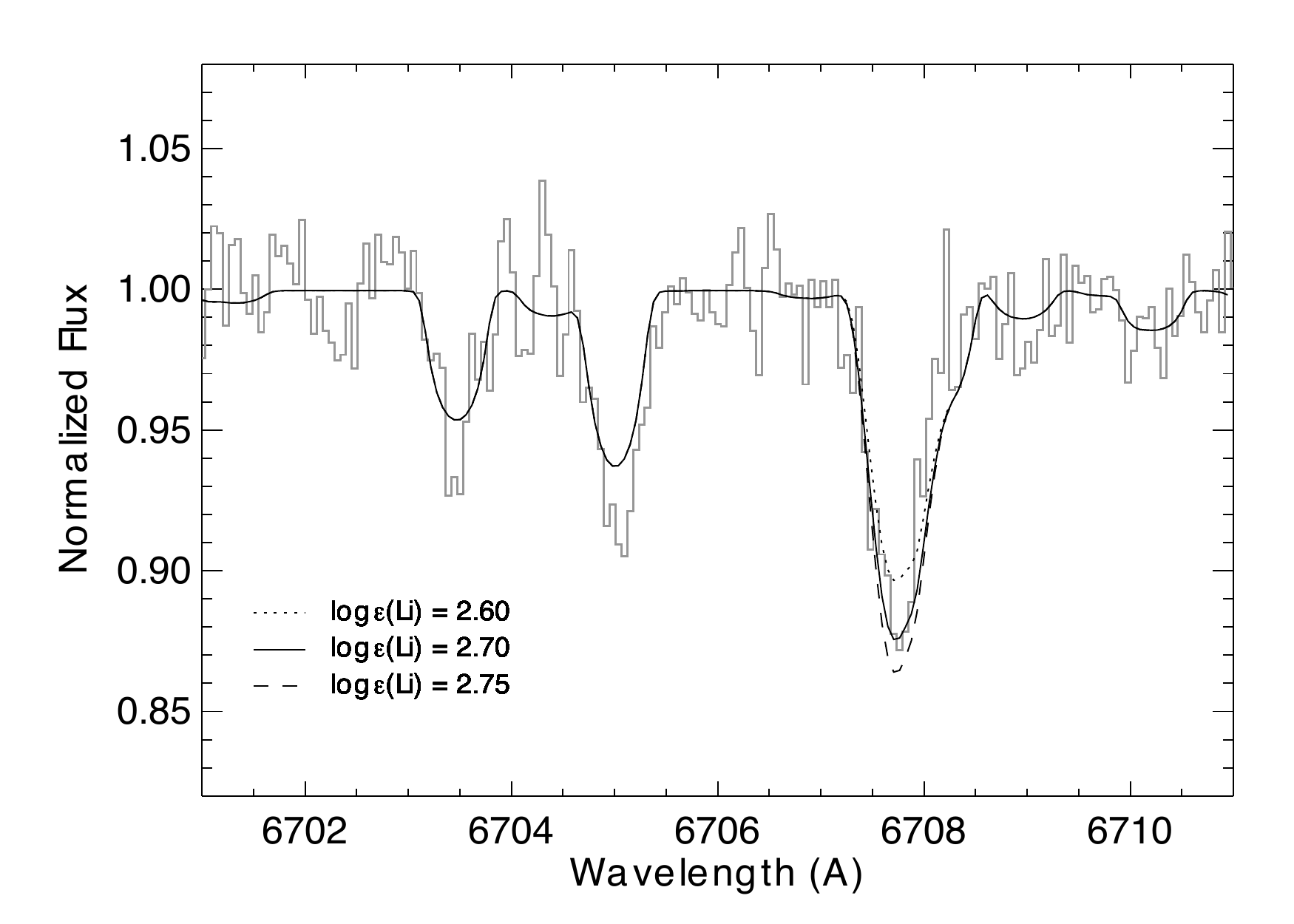}
\caption{A portion of the co-added observed spectrum of Qatar-7b host star containing the 
Li I 6707.8 \AA\ line plotted in gray. Overplotted are three synthetic spectra differing only by 
the Li abundance as depicted by lines of different style. The best fit to the Li | 6707.8 \AA\ 
line requires $\log{\epsilon}{\rm (Li)} = 2.70$.}
\label{fig:Li_line}
\end{figure}

The resonance line Li I $\lambda6707.8$ is by far the strongest Li feature in the optical
and is the source of most of the Li abundance estimates. We note, that because of the low 
ionization potential (5.39 eV), Li I lines can only be seen in stars with $\teff\ < 8500$ K. 
The host star of Qatar-7b displays a strong Li I $\lambda$6707.8 absorption line indicative 
of a young age. We have carried out a spectral synthesis of a 10 \AA\ region centered on 
the Li line using the SME package and the stellar atmospheric parameters obtained in 
Section \ref{subsec:SpecPars}. Figure\,\ref{fig:Li_line} shows a comparison of the 
co-added spectrum of Qatar-7b with three synthetic spectra that differ only in Li abundance. 
The best fit is obtained for Li abundance of $\log{\epsilon}{\rm (Li)} = 2.70$. By comparison 
with the Li abundance curves as a function of age for different \teff\ for open clusters 
(\citealt{SestitoRandlich05}) we estimate an age $\leq 1$ Gyr for Qatar-7b host star. 

To summarize, all three age estimates give values in the range 0.5-1.5 Gyr. A more precise 
age determination may be possible via astroseismology but would require a photometric 
data set of much better precision. In Table \ref{table:StarPars} we quote the age of 
Qatar-7 as 1 Gyr, which is representative of the range of estimates, and put the 
uncertainty conservatively at 0.5 Gyr.

\begin{table*}
\centering
\caption{Basic observational and spectroscopic parameters of Qatar-7b host star and 
photometry used for the SED fit.}
\label{table:StarPars}
\begin{tabular}{lllll}
\hline \hline
Parameter & Description & Value  & Source & Ref. \\ 
\hline
\multicolumn{2}{l}{Names} & & & \\
 & \multicolumn{3}{l}{3UC\,255-282464, 4UC\,636-128328}  & \\
 & \multicolumn{3}{l}{SDSS\,J235403.63+370118.5, 2MASS\,J23540364+3701185}   & \\
 & \multicolumn{3}{l}{GALEX\,J235403.7+370119, WISE\,J235403.63+370118.6}   & \\

 \multicolumn{2}{l}{Astrometry} & & & \\
$\alpha_{\mathrm{2000}}$ & RA (J2000) & $23^{\mathrm{h}}54^{\mathrm{m}}03.63^{\mathrm{s}}$ & GAIA &1 \\    
$\delta_{\mathrm{2000}}$  & DEC (J2000) & $+37^{\mathrm{o}}01^{\prime}18.57^{\prime\prime}$ & GAIA & 1 \\
$\pi$  & parallax, mas & 1.3884 $\pm$ 0.0501  & GAIA & 1 \\
  
\multicolumn{2}{l}{Photometry} & & & \\
NUV & GALEX NUV, mag & 18.18 $\pm$ 0.02 & GALEX & 2 \\
$B$ & Johnson $B$, mag & 13.69 $\pm$ 0.04 & APASS & 4 \\       
$V$ & Johnson $V$, mag & 13.03 $\pm$ 0.06 & APASS & 4 \\       
$u$ & Sloan $u$, mag & 14.62 $\pm$ 0.05 & PD10 & 3 \\       
$g$ & Sloan $g$, mag & 13.32 $\pm$ 0.04 & APASS & 4 \\       
$r$ & Sloan $r$, mag & 12.86 $\pm$ 0.08 & APASS & 4 \\    
$i$ & Sloan $i$, mag & 12.67 $\pm$ 0.08 & APASS & 4 \\     
$z$ & Sloan $z$, mag & 12.70 $\pm$ 0.01 & PD10 & 3 \\     
$J$  & 2MASS $J$, mag & 11.859 $\pm$ 0.023 & 2MASS & 4 \\
$H$  & 2MASS $H$, mag & 11.599 $\pm$ 0.021 & 2MASS & 4 \\
$K$  & 2MASS $K$, mag & 11.561 $\pm$ 0.020 & 2MASS & 4 \\
W1 & WISE1, mag & 11.481 $\pm$ 0.024 & WISE & 5 \\              
W2 & WISE2, mag & 11.504 $\pm$ 0.021 & WISE & 5\\                
W3 & WISE3, mag & 11.252 $\pm$ 0.142 & WISE & 5 \\               

\multicolumn{2}{l}{Spectroscopic parameters} & & & \\
       & Spectral type & F4V & this work & \\
\teff  & Effective temperature, K  & 6311$\pm$50 & this work & \\
\logg & Gravity, cgs  &  4.26$\pm$0.10 & this work & \\
\mh   & Metallicity      & 0.29$\pm$0.08 & this work & \\
$\gamma_{\rm abs}$ &Systemic velocity, \kms\ & $-4.7\pm0.1$ & this work & \\
\vrot & Rotational velocity, \kms & 13.8$\pm$0.5 & this work & \\
$P_{\rm rot}$ &Rotation period, days & $5.9\pm0.2$ & this work &\\
$\tau$ & Age, Gyr & $1.0\pm0.5$ & this work & \\
$A_{V}$ & Extinction, mag & $0.38\pm0.05$ & S\&F2011 & 8 \\
$d$ & Distance, pc & $720\pm26$ & GAIA & 1 \\

 \hline
\end{tabular}
\tablerefs{(1) GAIA DR2 \url{http://gea.esac.esa.int/archive/}, \\
	(2) GALEX \url{http://galex.stsci.edu/}, (3) \cite{Pickles2010}, \\
	(4) APASS9 \url{http://www.aavso.org/apass}, \\
	(5) 2MASS \url{http://irsa.ipac.caltech.edu/Missions/2mass.html}, \\
	(6) WISE \url{http://irsa.ipac.caltech.edu/Missions/wise.html}
	}
\end{table*}

\subsection{Orbital period determination} \label{subsec:OrbitalPeriod}
We used our follow-up light curves to determine the ephemeris of the system. From the 
eleven available light curves (as listed in Table\,\ref{table:ObsLog}) we did not include 
the 1st and 7-10th, because they are partial and cannot provide an \emph{independent} 
measurement of $T_{\rm C}$; the 5th, because it's simultaneous with the 4th, but has 
larger scatter; and the 11th, because of insufficient coverage of out-of-transit parts.

For the remaining four light curves, we first converted all time-stamps to the BJD$_{\rm TDB}$ 
timescale and proceeded to fit each one individually using {\sc EXOFASTv2} 
\citep{Eastman2017}. The resulting transit central times $T_{\rm C}$ and their uncertainties 
are listed in Table\,\ref{table:Tcen}. The best ephemeris was calculated by fitting a straight 
line through the measured $T_{\rm C}$ points. The resulting orbital ephemeris is

\begin{equation}
T_{\rm C} = 2458043.32075(16) + 2.032070(17)\,E 
\label{eq:OrbEphemeris}
\end{equation}

\noindent 
where $E$ is the number of cycles after the reference epoch, which we take to be the 
$y$-intercept of the linear fit, and the numbers in parentheses denote the uncertainty in 
the last two digits. In Equation (\ref{eq:OrbEphemeris}), the reference epoch and the period 
are in days on the BJD$_{TDB}$ scale, and their uncertainties correspond to 14\,s, and 1.5\,s, 
respectively. 

\begin{table}
\centering
\caption{Central times of Qatar-7b transits and their uncertainties. The last column corresponds 
to the first column of Table\,\ref{table:ObsLog}.}
\label{table:Tcen}
\begin{tabular}{ccc}
\hline\hline
Transit central time             & Cycle No. & Obs ID \\
BJD$_{TDB}$ - 2\,450\,000  &  & \\
\hline
8043.32067 $\pm$ 0.00023 &  0 & 2 \\
8051.44909 $\pm$ 0.00021 &  4 & 3 \\
8063.64153 $\pm$ 0.00033 & 10 & 4 \\
8112.41103 $\pm$ 0.00057 & 34 & 6 \\
\hline
\end{tabular}
\end{table}

\subsection{Excluding a false positive} \label{subsec:FPP}

Observational data described in the previous sections -- the shape and depth of the transit 
light curves plus the shape and amplitude of the RV curve -- can be described by a case of 
a planet orbiting a single star. Yet, there are other astrophysical scenarios that could mimic 
such a behaviour and include an eclipsing binary -- either a background or in a hierarchieal 
tripple system -- blended with the primary (and brighter) star. Here we present arguments 
to exclude such scenarios and to prove that the observed behaviour is not a false positive. 

Our first argument comes from the behaviour of the spectral line bisectors (see Table 
\ref{table:RV}). By now it is well understood (e.g., \citealt{Queloz01}, \citealt{Torres05}) that 
if measured radial velocities were a result of a blend with an eclipsing binary, the line 
bisectors would vary in phase with the photometric period and with an amplitude similar to 
that of the RV curve. As can be seen from bottom panel in Figure \ref{figure:q7bRV} in the 
case of Qatar-7 the line bisectors are essentially stationary throughout the photometric 
period. The weighted mean of the bisector distribution is $0\pm20$ \ms, while the RV 
amplitude is $K = 239\pm32$ \ms. This lends a strong argument that the observed RV 
pattern is a result of a gravitationally induced motion from a planet orbiting a single star.

A second argument supporting the planet scenario comes from the fact that transit light 
curves in all filters have equal depth after accounting for the limb darkening. In fact, our 
transit photometry is good enough to detect small differences in the depth due to the 
wavelength dependence of the limb darkening and the results are in full agreement with 
a central transit (depth decreases with wavelength) as expected given the estimated 
impact parameter $b$ (Table \ref{table:GlobalFit}). Equal transit depths at different 
wavelengths, however, do not preclude a scenario of a stellar companion of similar 
colors. For this reason, we consider this argument only as supportive of the planetary 
scenario, but note that it is in full agreement with the conclusion of this section. 

Finally, we compute the False Positive Probability (FPP) for all scenarios involving an eclipsing 
binary using the {\tt vespa} package (\citealt{Morton12}, \citealt{Morton15}). Originally, 
vespa was developed as a tool for statistical validation of planet candidates identified by 
the {\it Kepler} mission (e.g., \citealt{Morton16}) and its successor {\it K2} (e.g., 
\citealt{Crossfield16}, \citealt{Mayo18}).  Vespa calculates FPP via model selection among a 
number of physical scenarios that could potentially explain the transit signal: an unblended 
eclipsing binary (EB), a hierarchical-triple eclipsing binary (HEB), a background/foreground 
(chance-aligned) eclipsing binary (BEB), the double-period cases of all these eclipsing-binary 
scenarios, and finally the transiting planet scenario. The likelihoods and priors for each scenario 
are based on the shape of the transit signal, the star's location in the Galaxy, and single-, binary-, 
and triple-star model fits to the observed photometric and spectroscopic properties of the star, 
as well as the Gaia parallax. Additional constraints come from direct imaging from the MuSCAT2 
photometry (see Section \ref{subsec:FollowPhot}), which puts limits on the presence of 
potentially blending stars, and from a limit on the secondary transit (non-)detection estimated 
from our discovery curve --- phase folded and binned to the transit duration. We note also that 
all previous published uses of vespa have been on Kepler data, for which the photometric 
integration time per cadence is approximately 30 minutes --- for this work, where our photometry 
has 2-minute cadence, we update vespa to be able to use a custom integration time, which 
is important because longer integration times significantly smooth out the ingress/egress 
times, especially for short-period planets \citep{Kipping2010}. This updated version of vespa 
computes Qatar-7b to have about a 5\% probability of being caused by the EB scenario 
(a faint low-mass star eclipsing the primary), and negligible probability for any of the other 
false-positive scenarios. However, our radial velocity measurements completely exclude 
the presence of such a stellar-mass close companion; hence, we can remove the EB 
scenario from consideration by vespa, leaving completely negligible probability that 
the signal could be caused by a false positive. The only viable scenario is that of a planet 
orbiting a single star.

\subsection{Planetary system parameters} \label{subsec:EXOFAST}

We derived the physical parameters of the Qatar-7 planetary system by running a global 
fit of the available transit light curves (Figure \ref{fig:q7bTR}) and RV measurements 
(Table \ref{table:RV}) using {\sc EXOFASTv2} (\citealt{Eastman2017}, \citealt{Rodrigues2017}). 
The set of parameters used to initialize the fit consisted of the orbital ephemeris (Section 
\ref{subsec:OrbitalPeriod}), the stellar parameters (\teff, \logg, \mh) as determined through 
SPC, as well as the distance (via the GAIA parallax) and age (Section \ref{subsec:SpecPars}). 
In addition, we set the visual extinction to $A_{V}$=0.38, based on the Galactic dust reddening 
maps (\citealt{Extinction}). Gaussian priors were imposed on all these parameters, with the 
mean values and standard deviations quoted in Table \ref{table:StarPars}. Limb darkening 
coefficients (LDCs) were left free to vary, but also had a Gaussian prior imposed on them, with 
the mean value interpolated from the  \cite{LDcoeff} tables (for the given filter of each transit 
light curve used) and a standard deviation of 0.05. Finally, we kept the system eccentricity 
fixed to zero, as the orbit circularization timescale (using the \cite{Jackson08} equations) is 
calculated to be $\tau_{\rm circ} \leq 0.07$ Gyr, much lower than the estimated age of the 
host star. 

The results of the global fit are summarized in Table \ref{table:GlobalFit}. The Safronov 
number is not used in the current paper and is provided in Table \ref{table:GlobalFit} for 
completeness, as it may be useful for other studies.

\begin{table*}
\centering
\caption{Median values and 68\% confidence intervals. We assume 
	$R_{\odot}$=696342.0\,km, $M_{\odot}$=1.98855$\times 10^{30}$\,kg, 
	$R_{\rm J}$ = 69911.0\,km, $M_{\rm J}$=1.8986$\times 10^{27}$\,kg 
	and 1 AU=149597870.7 km.}
\label{table:GlobalFit}
\begin{tabular}{lcc}
\hline
Parameter & Units & Qatar-7b \\
\hline
\multicolumn{2}{l}{Stellar Parameters:} &  \\
    ~~~$M_{*}$\dotfill     &Mass (\msun)\dotfill  & $1.409\pm0.026$ \\
    ~~~$R_{*}$\dotfill      &Radius (\rsun)\dotfill & $1.564\pm0.021$ \\
    ~~~$L_{*}$\dotfill       &Luminosity (\lsun)\dotfill & $3.66\pm0.13$ \\
    ~~~$\rho_*$\dotfill     &Density (g/cm$^{3}$)\dotfill & $0.521\pm0.016$ \\
    ~~~$\log(g_*)$\dotfill &Surface gravity (cgs)\dotfill & $4.196\pm0.029$ \\
    ~~~$\teff$\dotfill        &Effective temperature (K)\dotfill & $6387\pm38$  \\
    ~~~$\feh$\dotfill        &Metallicity\dotfill & $0.276\pm0.071$ \\
    ~~~$\tau_{\rm YY}$\dotfill   &Age (Gyr)\dotfill & $1.69\pm0.25$   \\
    ~~~$A_{V}$\dotfill    & Extinction (mag)\dotfill & $0.338\pm0.055$ \\
    ~~~$d$\dotfill &Distance (pc)\dotfill & $725\pm10$   \\
\multicolumn{2}{l}{Planetary Parameters:} &  \\
    ~~~$P$\dotfill   &Period (days)\dotfill & $2.032046\pm0.0000097$ \\
    ~~~$a$\dotfill   &Semi-major axis (AU)\dotfill & $0.0352\pm0.0002$ \\
    ~~~$M_{P}$\dotfill &Mass (\mj)\dotfill & $1.88\pm0.25$ \\
    ~~~$R_{P}$\dotfill &Radius (\rj)\dotfill & $1.70\pm0.03$ \\
    ~~~$\rho_{P}$\dotfill &Density (g/cm$^{3}$)\dotfill & $0.502\pm0.071$ \\
    ~~~$\log(g_{P})$\dotfill &Surface gravity\dotfill & $3.224\pm0.066$ \\
    ~~~$T_{eq}$\dotfill &Equilibrium Temperature (K)\dotfill & $2053\pm15$ \\
    ~~~$\Theta$\dotfill &Safronov Number\dotfill & $0.056\pm0.008$ \\
\multicolumn{2}{l}{RV Parameters:} &  \\
    ~~~$K$\dotfill &RV semi-amplitude (m/s)\dotfill & $239\pm32$\\
    ~~~$\gamma_{\rm rel}$\dotfill &Systemic velocity (m/s)\dotfill & $106\pm35$ \\
    ~~~$e$\dotfill &Eccentricity (fixed)\dotfill & $0$\\
\multicolumn{2}{l}{Primary Transit Parameters:} &  \\
    ~~~$R_{P}/R_{\star}$\dotfill &Radius of planet in stellar radii\dotfill & $0.110\pm0.001$ \\
    ~~~$a/R_{\star}$\dotfill &Semi-major axis in stellar radii\dotfill & $4.84\pm0.05$\\
    ~~~$i$\dotfill &Inclination (degrees)\dotfill & $89.0\pm1.0$\\
    ~~~$b$\dotfill &Impact Parameter\dotfill & $0.08\pm0.08$ \\
    ~~~$T_{14}$\dotfill &Total duration (days)\dotfill & $0.1491\pm0.0009$\\
\hline
\end{tabular}
\end{table*}
\noindent

\section{Discussion and Conclusions} \label{sec:conclusions}

In this article, we present Qatar-7b, a newly identified very hot Jupiter orbiting close to 
a moderately fast rotating F-star. From a global fit to the available photometric and spectroscopic 
follow-up observations we measure the planet's mass 1.88\mj\, and radius 1.70\rj, which
puts Qatar-7b in the top 6\% of the distribution of known exoplanets by size. With an 
orbital period $P = 2.032$ d, the planet is located only 4.84 \rstar\ from the host star 
and is heated to a very high equilibrium temperature $\teq \approx 2100$ K putting it 
also within the top $\sim6\%$ of the hotest known exoplanets. 

Close-in transiting giant planets offer the best opportunities to look for dynamic interactions 
that may provide important information related to planetary system architecture and evolution. 
These include studies of the spin/orbit alignment through Rossiter-McLaughlin (RM) effect 
(\citealt{Rossiter1924}, \citealt{McLaughlin1924}) or Doppler Tomorgraphy (see \citealt{Zhou16a}, 
\citealt{Zhou16b}, \citealt{Zhou17}) and the search for presence of additional bodies in the 
system that may be revealed through transit timing and/or transit duration variations 
(\citealt{Kipping2010}). In addition, close-in transiting planets are often inflated due to the 
proximity and intense irradiation by the parent star, and are the best targets  to study the 
atmospheric composition in extrasolar planets (see \citealt{Murgas2017} and references 
therein). 

\begin{figure}
\centering
\includegraphics[width=8.8cm]{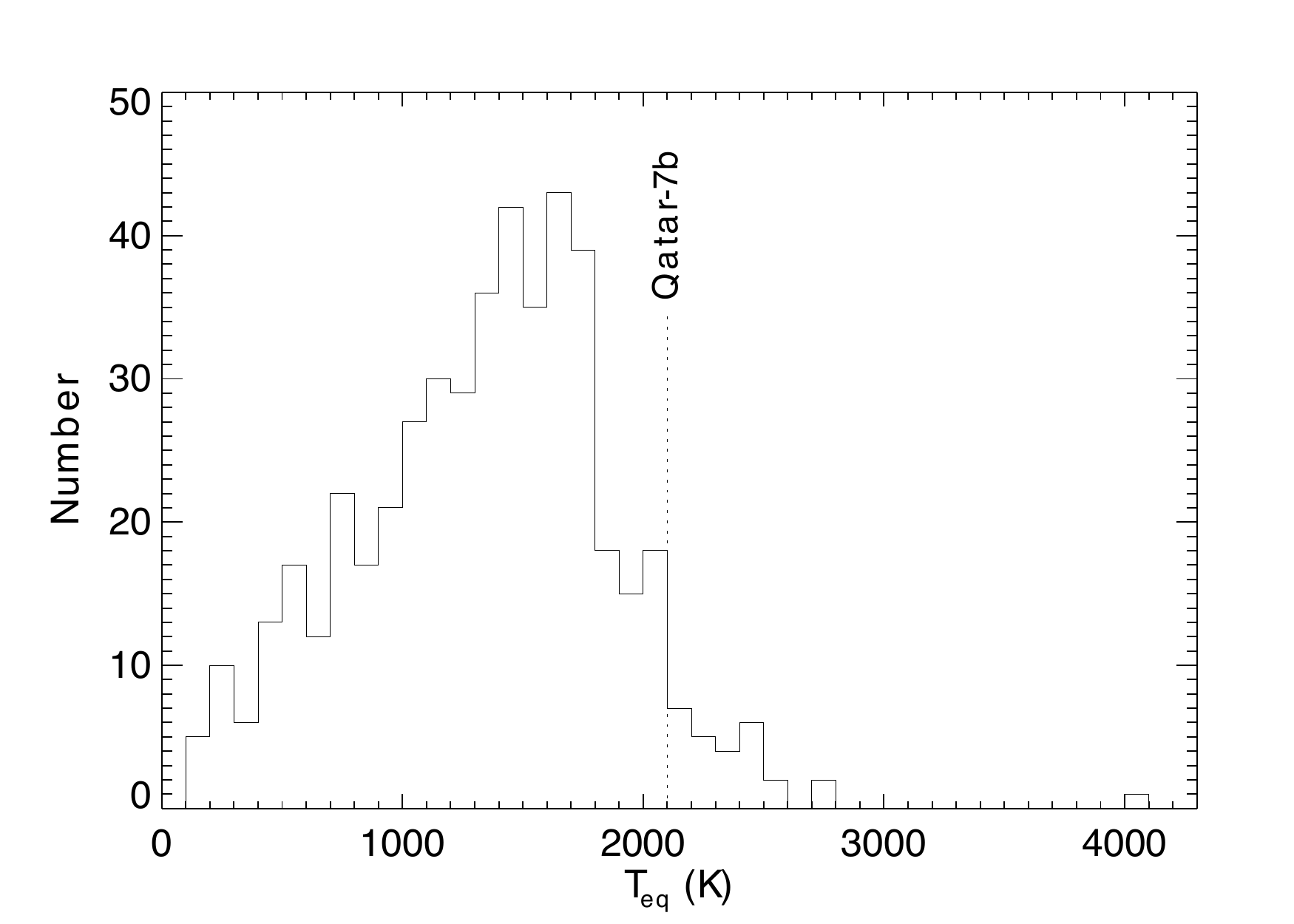}
\caption{Histogram of the equlibrium temperature of exoplanets.}
\label{fig:TeqHist}
\end{figure}

\begin{figure}
\centering
\includegraphics[width=8.8cm]{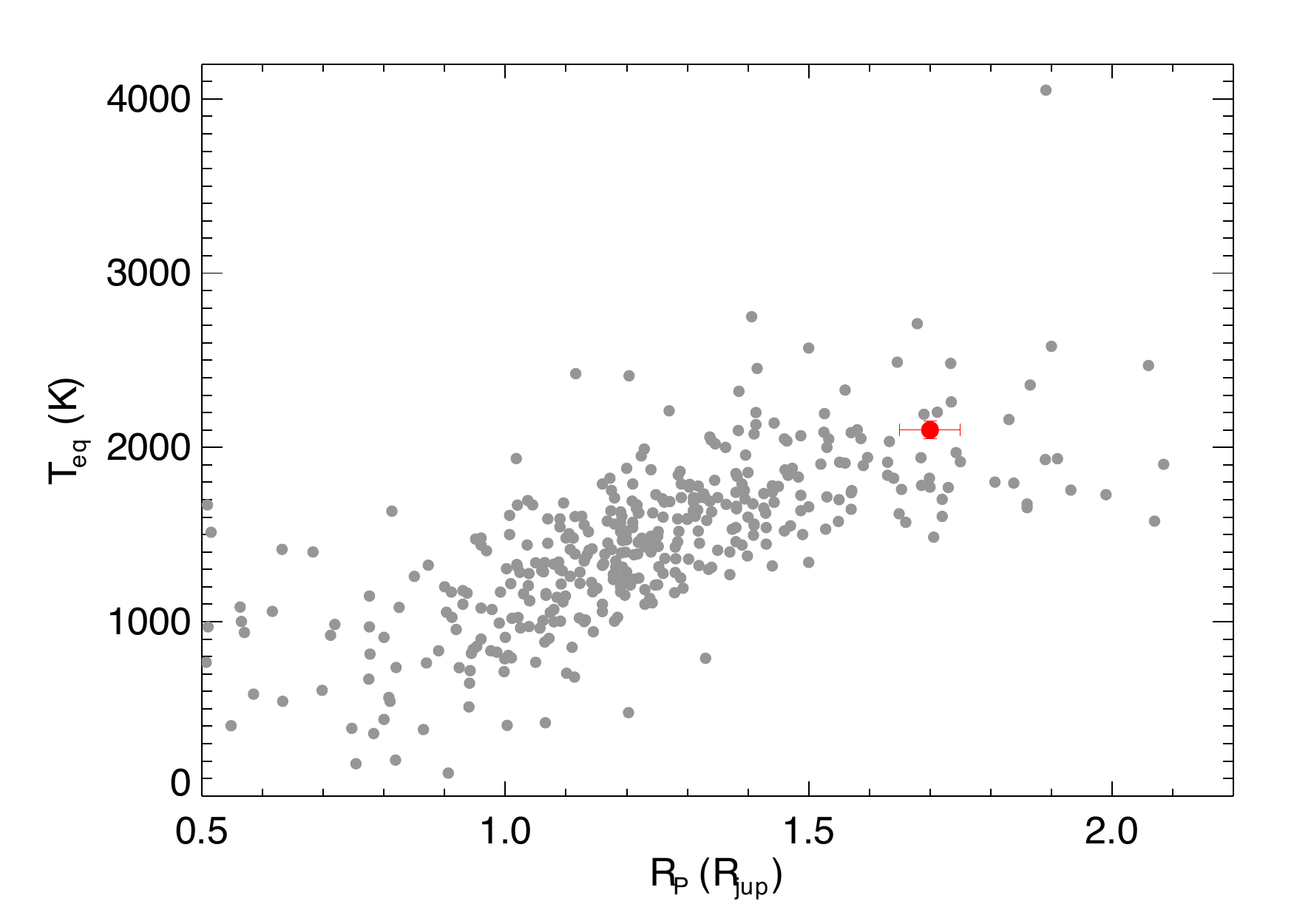}
\caption{The planet radius--equlibrium temperature, \rpl/\teq, relation for exoplanets. The 
position of Qatar-7b is plotted with larger red symbol with error bars.}
\label{fig:RpTeq}
\end{figure}

By its characteristics, Qatar-7b is an excellent target for dynamical studies. The host is a 
fast rotating star, $\vsini \approx 15\,\kms$ and combined with the relevant stellar and 
planetar parameters the predicted semi-amplitude of the RM effect is $\sim$200 \ms. 
With modern precision RV instruments the shape of the curve is expected to be securely 
measurable and allow for a good estimate of the stellar obliquity in the system. And last 
but not least, Qatar-7b is also an excellent target for atmospheric studies via transmission 
spectroscopy due to its significantly inflated planetary radius and very high equlibrium 
temperature. In Figures \ref{fig:TeqHist} and \ref{fig:RpTeq}, using data for the well studied 
exoplanets from TEPcat\footnote{The Transiting Extrasolar Planet Catalog (TEPcat) is 
available at http://www.astro.keele.uk/jkt/tepcat}, we show that Qatar-7b falls within the 
top $\sim6\%$ of the hottest exoplanets known so far and occupies a place close to the 
extreme end of the $\rpl/\teq$ relation. 

In their study for HAT-P-65b and HAT-P-66b, \cite{Hartman16} find that large, inflated hot 
Jupiters ($\rpl\,>1.5\rj$) are preferentially found around older and/or slightly evolved stars. 
Defining the fractional age of the host star as 
$\tau = (t_{\rm cur}-200\,\mathrm{Myr})/(t_{\rm tot}-200\,\rm{Myr})$, with $t_{\rm cur}$ 
and $t_{\rm tot}$ being the current and total lifetime of the star respectively, they find that 
the majority of stars hosting large hot Jupiters have $\tau > 0.5$, i.e., that they have spent 
more than half of their main-sequence lifetime. With a radius of \rpl\ = 1.70\rj, Qatar-7b 
comfortably falls in this category, but seems not to conform with the aforementioned trend, 
as the fractional age of the host star is calculated to be $\tau \approx 0.3$, for a total 
lifetime of $t_{\rm tot} \approx 3.5$ Gyr, as tabulated from the YY isochrones for a star of 
the same mass and metallicity. 

\section*{Acknowledgements}

This publication is supported by NPRP grant no. X-019-1-006 from the Qatar National 
Research Fund (a member of Qatar Foundation). The statements made herein are solely 
the responsibility of the authors. The Nanshan 1m telescope of XAO is supported by the CAS 
"Light of West China" program (XBBS-2014-25,2015-XBQN-A-02), and the Youth Innovation 
Promotion Association CAS (2014050). This article is partly based on observations made in the 
Observatorios de Canarias del IAC with the MUSCAT2 instrument on the Carlos Sanchez 
telescope operated on the island of Tenerife by the IAC in the Observatorio del Teide. This 
work is partly financed by the Spanish Ministry of Economics and Competitiveness through 
grants ESP2013-48391-C4-2-R. This work has made use of data from the European Space 
Agency (ESA) mission {\it Gaia} (\href{https://www.cosmos.esa.int/gaia}{https://www.cosmos.esa.int/gaia}), 
processed by the {\it Gaia} Data Processing and Analysis Consortium (DPAC,
\href{https://www.cosmos.esa.int/web/gaia/dpac/consortium}{https://www.cosmos.esa.int/web/gaia/dpac/consortium}). 
Funding for the DPAC has been provided by national institutions, in particular the 
institutions participating in the {\it Gaia} Multilateral Agreement. We also acknowledge 
support from JSPS KAKENHI Grant Number 16K13791.




\end{document}